# Radiation forces and confinement of neutral particles into the pulse envelope. New regime of collision ionization.

L. M. Kovachev

Academician Emil Djakov Institute of Electronics, Bulgarian Academy of Sciences, 72 Tzarigradsko choussee, 1784 Sofia, Bulgaria

e-mail: lubomirkovach@yahoo.com

**Abstract**. For a long time, transverse optical forces with continuous-wave laser beams have been used for non-contact manipulation of micro-scale objects. By replacing the continuous-wave lasers with pulse-mode lasers many novel physical effects related with the influence of Poynting vector on the longitudinal part of the optical force can be observed. The following question arises: What is the impact of this longitudinal addition to the optical force, connected with the flow of energy, on the ensemble of thousand particles in continuous media? The aim of this work is to find analytical expressions of the radiation force and potential densities arising from a laser pulse propagating in a dielectric media. This allows us to find an effective averaged longitudinal real force at the level of the laser pulse's spot. The obtained force is proportional to the initial pulse energy and inversely proportional to its time duration. In the femtosecond region the force becomes strong enough to confine neutral particles into the pulse envelope and translate them with the group velocity in gases. Additionally, if the trapped particles into the pulse's spot are with high density, the probability of collision with the free atoms and molecules in air become significant. The collision energies are in the range of 12 - 24 GeV and high enough to ionise the neutral atoms. Thus, a new type of collision ionisation can be observed, when powerful femtosecond pulses propagate in gaseous media.

## 1. Introduction

Particles can be trapped by laser pulses working in CW regime, as demonstrated by Ashkin [1, 2]. The analytical expression of the radiation force on one individual particle is obtained in dipole approximation and as it is well known, is proportional to the transverse gradient of the square of the electrical field. In case where the CW regime is replaced with optical pulse propagation, additional terms associated with the Poynting vector and the flow of energy appear [3, 4]. The forces are connected with the polarization of atoms and molecules, and depending of the initial power, the nonlinear polarization can play a significant role [5]. The question, what kind of radiation forces applied to ensemble of neutrals with linear and nonlinear polarization, from a laser pulse in continuous dielectric media is still open. In this paper is obtained and discussed the analytical expressions of the longitudinal radiation force density, and after integration – the effective real force at the level of the pulse width in approximation of first order of dispersion. The influence of the nonlinear polarization on the radiation force is considered. The obtained force density is proportional to the first derivative of the pulse time envelope, while the real force at the level of the pulse width is inversely proportional to the pulse time duration. This is the reason the force to vanish in CW regime, while in the femtosecond region leads to trapping of particles into the pulse envelope. When the trapped particles into the pulse's spot are with high density, the probability of collision with the free atoms and molecules in air becomes significant.

**2. Nonlinear polarization and radiation forces**
The conditions for Rayleigh scattering are fulfilled in the case when the diameter of the particle is significantly smaller than the wavelength of the electric field, and the particle can be considered as a point dipole in the field of an electromagnetic pulse [3]. The force on a particle of charge $q$ moving at velocity $v$ though an electromagnetic field is the Lorentz force law given by

$$\vec{F} = q\left(\vec{E} + \frac{v}{c} \times \vec{B}\right). \tag{1}$$

Assuming that the effective centre of the electron cloud of an individual atom or molecule is located at the point $x_1$ and the effective centre of the positive nucleus is at the point $x_2$, the Lorentz force applied to the two opposite charges by an electromagnetic pulse can be presented as follow:

$$\vec{F_1} = q\left(\vec{E}(x_1, y, z, t) + \frac{1}{c}\frac{dx_1}{dt} \times \vec{B}(x_1, y, z, t)\right), \tag{2}$$

$$\vec{F_2} = -q\left(\vec{E}(x_2, y, z, t) + \frac{1}{c}\frac{dx_2}{dt} \times \vec{B}(x_2, y, z, t)\right). \tag{3}$$

The expression for the polarization of a dipole is:

$$\vec{p} = q\vec{d} = q(\vec{x}_1 - \vec{x}_2), \tag{4}$$

where $\vec{d}$ is the distance between the two charges. In the case of an infinitesimal displacement, when $\vec{x}_1 - \vec{x}_2$ is negligible with respect to the wavelength, the Lorentz force on a dipole can be obtained as:

$$\vec{F} = \vec{F_1} - \vec{F_2} = q\left(\vec{E}(x_1, y, z, t) - \vec{E}(x_2, y, z, t) + \frac{1}{c}\left(\frac{d\vec{x}_1}{dt} \times \vec{B}(x_1, y, z, t) - \frac{d(\vec{x}_2)}{dt} \times \vec{B}(x_2, y, z, t)\right)\right) \cong$$
$$q\left((\vec{x}_1 - \vec{x}_2) \cdot \nabla \vec{E} + \frac{1}{c}\frac{d(\vec{x}_1 - \vec{x}_2)}{dt} \times \vec{B}\right). \tag{5}$$

From Equation (4) and Equation (5) is obtained:

$$\vec{F} = (\vec{p} \cdot \nabla)\vec{E}(x, y, z, t) + \frac{1}{c}\frac{d\vec{p}}{dt} \times \vec{B}(x, y, z, t). \tag{6}$$

Thus, as pointed in [3], the polarization force is evidently a Lorentz type. The mutual action of the linear and nonlinear polarization in a gaseous media (air) is investigated in the current research. In this case the polarization of one individual atom or molecule is

$$\vec{p} = \alpha \vec{E} + \gamma^{(3)} |\vec{E}|^2 \vec{E}, \tag{7}$$

where $\alpha$ and $\gamma^{(3)}$ are the linear and nonlinear polarizability of one individual atom or molecule correspondingly. Substituting Equation (7) in Equation (6) leads to the following formula expression:

$$\vec{F} = \left(\alpha + \gamma^{(3)} |\vec{E}|^2\right)(\vec{E} \cdot \nabla)\vec{E} + \frac{1}{c} \frac{d\left(\alpha + \gamma^{(3)} |\vec{E}|^2 \vec{E}\right)}{dt} \times \vec{B}. \tag{8}$$

To obtain density force of ensemble of particles, the optical force of an individual atom is multiplied by the number of atoms per volume $N$ and the local field correction is neglected as small one for gases. Then, Equation (8) transforms to

$$\vec{f} = N\vec{F} = \left(\chi^{(1)} + \chi^{(3)} |\vec{E}|^2\right)(\vec{E} \cdot \nabla)\vec{E} + \frac{1}{c} \frac{d\left(\chi^{(1)} + \chi^{(3)} |\vec{E}|^2\right)\vec{E}}{dt} \times \vec{B}, \tag{9}$$

where $\vec{f}$ is the force density of the media, $\chi^{(1)} = N\alpha$ and $\chi^{(3)} = N\gamma^{(3)}$ are the linear and nonlinear susceptibility of the media. Applying the relation

$$(\vec{E} \cdot \nabla)\vec{E} = \frac{1}{2}\nabla(\vec{E}^2) - E \times (\nabla \times E), \tag{10}$$

and the first Maxwell equation $\nabla \times \vec{E} = -\frac{1}{c} \frac{\partial \vec{B}}{\partial t}$, one can obtain

$$\vec{f} = \left(\chi^{(1)} + \chi^{(3)} |\vec{E}|^2\right)\left[\frac{1}{2}\nabla(\vec{E}^2) + \frac{1}{c} \vec{E} \times \frac{d}{dt}\vec{B}\right] + \left[\frac{1}{c} \frac{d\left(\chi^{(1)} + \chi^{(3)} |\vec{E}|^2\right)\vec{E}}{dt} \times \vec{B}\right]. \tag{11}$$

Introducing the Poynting vector by $\vec{S} = \frac{c}{4\pi} \vec{E} \times \vec{B}$ ($\mu_0 = 1$) the expression for nonlinear radiation force from Equation (11) can be written as

$$\vec{f} = \frac{1}{2}\left(\chi^{(1)} + \chi^{(3)} |\vec{E}|^2\right)\nabla(\vec{E}^2) + \frac{4\pi}{c^2} \frac{d\left[\left(\chi^{(1)} + \chi^{(3)} |\vec{E}|^2\right)\vec{S}\right]}{dt}. \tag{12}$$

The first term in Equation (12) presents the extended Ashkin's gradient term, including additionally longitudinal and nonlinear parts to the well-known transverse gradient ones. The second term is a nonlinear generalization of the longitudinal linear radiation force term investigated in [3, 4].

### 3. Nonlinear radiation force of a linearly polarized laser pulse
The electrical and magnetic field of a laser pulse are presented by amplitude and carrying frequency

$$\vec{E} = \left(\vec{A}(x,y,z,t)\exp(i(\omega_0 t - k_0 z)) + c.c.\right)/2, \tag{13}$$

where $\vec{A}(x,y,z,t)$ is the complex amplitude of the electrical field, $\omega_0$ is the carrying frequency and $k_0$ is the carrying wave number. The standard linear polarized initial laser pulse with component $\vec{A} = (A_x, 0, 0)$ is considered and applied. Let's initially investigate the longitudinal part of the Ashkin's gradient force for such laser pulse

$$f_z^{Ash} = \frac{1}{2}\left(\chi^{(1)} + \chi^{(3)}|\vec{E}|^2\right)\frac{\partial}{\partial z}(\vec{E}^2) = \left(\chi^{(1)} + \gamma^{(3)}|A_x|^2\right)\left(A_x \frac{\partial A_x}{\partial z} - ik_0 A_x\right)\exp(2i(\omega_0 t - k_0 z)) + c.c. \tag{14}$$

The difference from the transverse components of the Ashkin's force is that the longitudinal gradient force oscillate and its averaged value on the propagation distance vanish

$$\langle f_z^{Ash} \rangle = 0. \tag{15}$$

The next step is to investigate the second term in Equation (12), connected with the flow of energy of a laser pulse and the Poynting vector

$$\vec{f}^P = \frac{4\pi}{c^2}\frac{d\left[\left(\chi^{(1)} + \chi^{(3)}|\vec{E}|^2\right)\vec{S}\right]}{dt}. \tag{16}$$

To solve the issue, at this point the differences between the atomic and optical scales is considered. The atoms and molecules can be characterized by their atom (molecular) response of order of $\tau_0 \approx 2-3$ fs. During this time the laser pulse propagates at distance of $z_{resp} = \tau_0 v_{gr} \approx 0.5-1.$ μm. The optical scale is characterized by diffraction length and for a typical laser pulse varies widely in the range of $z_{diff} = k_0 d_0^2 \approx 15-1500$ cm. Since $z_{diff} >> z_{resp}$ is always at one diffraction length there are thousand oscillations of the atom dipole. Thus, the spot of the pulse does not change significantly at distances less than one diffraction length. There is also possibility to enlarge this distance up to few diffraction lengths, using fs pulses in diffraction free regime [6] and in nonlinear wave-guiding regime with power slightly higher than the critical for self-focusing [7-9] propagates in air. In these regimes the transfer shape (the spot) of the pulse does not change of few diffraction lengths, and the flow of energy is in a surface orthogonal to the propagation direction. The divergence of $\vec{S}$ in this regime is equal to $\nabla \cdot \vec{S} = \vec{z}\,\partial S_z/\partial z$. This allows the use of the Poynting vector approximation in the form $\vec{S} = (0, 0, S_z)$. In our previous work [4] the equation of flow of energy is used to solve the problem. In this paper is considered that in wave-guiding regime and the Pointing vector from Equation (16) can be expressed by the intensity profile $\vec{S} = S_z \vec{z} = \left[\frac{n_0 c}{2\pi}|A_x|^2\right]\vec{z}$, as well as the modulus of the square of electrical field is $|\vec{E}|^2 = |A_x|^2$. Finally, the linear and nonlinear longitudinal radiation force of a linearly polarized laser pulse in wave-guiding regime is given by the form:

$$f_z^P = \frac{2n_0\chi^{(1)}}{c}\frac{d|A_x|^2}{dt} + \frac{2n_0\chi^{(3)}}{c}\frac{d(|A_x|^4)}{dt}. \tag{17}$$

**4. Longitudinal radiation force density and potential in approximation to the first order dispersion of a Gaussian pulse**

The difference between the atomic and optical scales, as well as the diffraction-free and nonlinear wave-guiding regimes gives the possibility to simplify the vector Equation (16) to a scalar Equation (17), representing the longitudinal radiation force of a laser pulse, associated with the Poynting vector. The solution of initially Gaussian pulse in approximation of the first order of dispersion in wave-guiding regime or in the frame of spatio-temporal paraxial optics, at distances smaller than the diffraction and dispersion lengths is

$$A(t, x, y, z) = A_0 \exp\left(-\frac{x^2+y^2}{2d_0^2} - \frac{(z-v_{gr}t)^2}{2z_0^2}\right), \tag{18}$$

where $d_0^2$ is the pulse spot, and $z_0 = v_{gr}t_0$ is its longitudinal length. Substituting Equation (18) in Equation (17) and differentiating on time, one can obtain the analytical expression of longitudinal force density of a laser pulse associated with the Poynting vector

$$f_z^P = -\frac{4n_0 v_{gr}}{c}\frac{A_0^2}{z_0^2}\left[\begin{array}{l}\chi^{(1)}\exp\left(-\dfrac{x^2+y^2}{d_0^2}\right)\exp\left(-\dfrac{(z-v_{gr}t)^2}{z_0^2}\right)(z-v_{gr}t) + \\ 2\chi^{(3)}A_0^2\exp\left(-\dfrac{2(x^2+y^2)}{d_0^2}\right)\exp\left(-\dfrac{2(z-v_{gr}t)^2}{z_0^2}\right)(z-v_{gr}t)\end{array}\right]. \tag{19}$$

The first term in the brackets of Equation (19) corresponds to the linear polarization of the media, while the second one is associated with the nonlinear polarization. It can be seen that the longitudinal radiation force is connected with the pulse envelope and propagates along with the group velocity. This allows the real 3D shape of the radiation force to be presented in Cartesian ($x$, $y$, $z' = z - v_{gr}t$) coordinates. The force density is gradient ones in $z'$ direction and a potential density can be introduced by

$$U(x, y, z) = \int_{-\infty}^{z'} f_{z'}^P dz'. \tag{20}$$

Integrating Equation (19) in $z'$ direction we obtain

$$U_z^P = -\frac{2n_0 v_{gr}}{c} A_0^2 \left[ \begin{array}{l} \chi^{(1)} \exp\left(-\frac{x^2+y^2}{d_0^2}\right) \exp\left(-\frac{z'^2}{z_0^2}\right) + \\ \chi^{(3)} A_0^2 \exp\left(-\frac{2(x^2+y^2)}{d_0^2}\right) \exp\left(-\frac{2z'^2}{z_0^2}\right) \end{array} \right]. \quad (21)$$

The expressions for radiation forces (19) and potentials (21) are written in standard Gaussian units. In the experiments the conventional system of units is used. This is the reason to rewrite the equation of radiation forces (19) and potentials (21) in conventional system of units after replacing of square modulus of amplitude function with the intensity $I_0$, and nonlinear susceptibility by nonlinear refractive index $n_2$ [10]

$$|I_0|^2 \left[\frac{J}{cm^2 \sec}\right] = \frac{cn_0}{2\pi} \left\langle |A_0|^2 \left[\frac{J}{cm^3}\right] \right\rangle = 10^7 \frac{cn_0}{2\pi} \left\langle |A_0|^2 \left[\frac{erg}{cm^3}\right] \right\rangle, \quad (22)$$

$$n_2 \left[\frac{cm^2 \sec}{J}\right] = 10^{-7} \frac{12\pi^2}{n_0^2 c} \left\langle \chi^{(3)} \left[\frac{cm^3}{erg}\right] \right\rangle. \quad (23)$$

By multiplying Equation (22) and Equation (23) the dimensionless nonlinearity in both systems, conventional and Gausian is compared

$$n_2 I_{conv} = \frac{6\pi}{n_0} \chi^3 |A_0|^2_{Gaus}. \quad (24)$$

In the radiation force (19) and potential (21) equations, the influence of the linear and the nonlinear part for different intensities of the laser pulses can be compared. The nonlinear part becomes significant when $\chi^{(1)} \approx \chi^3 |A_0|^2$. In air this corresponds to intensity equal to $I_0 \approx 1.1 \times 10^{16}$ $[W/cm^2]$, which is of the order of ionization breakdown of the media. The self-guiding regime of femtosecond pulses with power less and nearly above the critical for self-focusing is investigated, where the intensities reach the values of $I_0^{self-guiding} \leq 1. \times 10^{12}$ $[W/cm^2]$.

The nonlinearity does not add significant amounts to the force and potential and because of this in diffraction-free and self-guiding regime it is ignored. Now, rewriting Equations (19) and (21) in expressions for nonlinear refractive index and intensity by using the relations in Equations (22)-(24), and neglecting the nonlinear parts as small ones one obtains optical forces and potentials for pulses in wave-guiding regime of propagation:

$$f_z^P = -\frac{8\pi \chi^{(1)} v_{gr}}{c^2} \frac{I_0}{z_0} \exp\left(-\frac{x^2+y^2}{d_0^2}\right) \exp\left(-\frac{z'^2}{z_0^2}\right) \frac{z'}{z_0} \quad (23)$$

$$U_z^P = -\frac{2\pi \chi^{(1)} v_{gr}}{c^2} I_0 \exp\left(-\frac{x^2+y^2}{d_0^2}\right) \exp\left(-\frac{z'^2}{z_0^2}\right). \quad (24)$$

The 3D image of the force density is plotted and presented in Figure 1. The pulse front attracts the ensemble of particles to the center of the pulse while the back side pushes them again to the center.

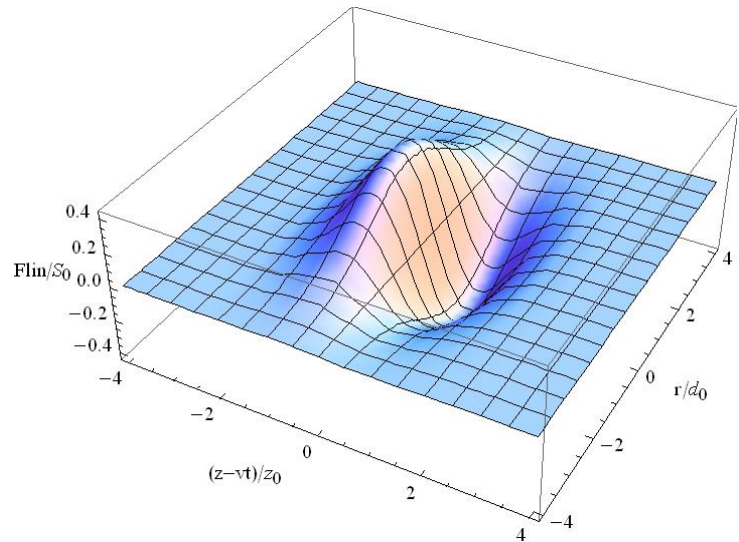

Fig. 1. Graphics of the PM longitudinal force density of a laser pulse. The pulse front attracts the ensemble of particles to the center of the pulse, and the back side pushes them again to the center.

The graph of the potential density is plotted in Figure 2. The Gaussian shape of the pulse plays the role of an attractive potential.

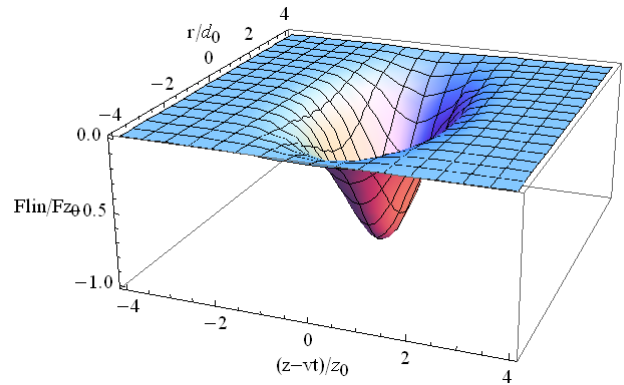

Fig. 2. Graphics of the potential density of a Gaussian laser pulse. The shape of the pulse, moving with the group velocity, plays the role of an attractive potential.

## 5. Effective averaged force and potential at the level of the pulse width

To obtain real measurable forces from the density ones, the formulae for the force density Equation (19) and the potential density Equation (21) is essential to be integrated over the whole space. As a result, after integration, two additional constants will be obtained - the square of the diameter of the pulse $d_0^2$ and the longitudinal size $z_0 = v_{gr} t_0$. In this way an effective real force at the level of the pulse spot expressed in conventional units is

$$\langle F \rangle_z^P = -\frac{v_{gr}^2}{c^2}\left[\pi^2 \chi^{(1)} + 6\pi n_2 I_0\right]\frac{E_0}{z_0} = -\frac{v_{gr}}{c^2}\left[\pi^2 \chi^{(1)} + 6\pi n_2 I_0\right]\frac{E_0}{t_0}, \qquad (25)$$

while for the potential is obtained

$$\langle U \rangle_z^P = -\frac{v_{gr}^2}{c^2}\left[\pi^2 \chi^{(1)} + 6\pi n_2 I_0\right] E_0, \qquad (26)$$

where $E_0$ is the initial energy of the laser pulse. The optical force, as gradient ones, is inversely proportional to the longitudinal length or temporal duration of the laser pulse. The long pulses are with relatively small gradient in respect to femtosecond ones and thus, this force usually is negligable. Comparing the expressions in the braskets one can see again that the nonlinearity does not add significant valuses for intensities up to $I_0 \approx 10^{12-13}\ W/cm^2$. That is why in the present examples we calculate the linear part of the force and potential only. Here a question arises - How deep is the radiation potential in air? Let's compare it to the Boltzmann energy of free particles at room temperature $T = 300\ K$. The value than is

$$U_B = k_B T = 4.14 \times 10^{-21}\ [J]. \qquad (27)$$

In our example is used $t_0$=100 fs laser pulse with initial energy in range of $E_0^{laser} \cong 1\ mJ$ and the potential is

$$U_z^{eff} \cong 3. \times 10^{-9}\ [J], \qquad (28)$$

which is twelve orders of magnitude greater than the Boltzmann energy. The Boltzmann factor is very small

$$R = \exp\left(-\frac{U_z^{eff}}{k_B T}\right) \ll 1. \qquad (29)$$

When the Boltzmann factor is very small, this gives us one quantitative prediction for cooling and confinement of neutral particles into the pulse envelope. All calculations above confirm that a confinement of neutral particles into a femtosecond pulse by the longitudinal radiation force and potential is possible.

### 6. New regime of collision ionisation

The investigations in the previous section point that a confinement into the pulse envelope of *neutral* atoms or molecules (nitrogen's N and oxygen's $O_2$) in air can be realized by femtosecond optical pulses, propagating in diffraction-free regime (with power values less than the critical one for self-focusing [6]) or in nonlinear wave-guiding regime [7-9]. These regimes correspond to initial pulse energy in the range from 10 nJ up to 1-2 mJ (pulse duration around $t_0 \approx 100\ fs$ and width $d_0 \approx 1-10\ mm$). Let's now assume that the trapped into the pulse envelope particles move along with the group velocity. Then, the kinetic energy of translation of one nitrogen atom can be given as

$$E_N^{kinetic} = \frac{m_N v_{gr}^2}{2} \cong 12\ [GeV], \qquad (30)$$

while the energy of translation of one oxygen molecule is in the order of

$$E_{O_2}^{kinetic} = \frac{m_{o_2} v_{gr}^2}{2} \cong 24 \quad [GeV]. \tag{31}$$

The probability of collision with the free atoms and molecules in air become significant if we realize the high density of trapped particles into the pulse's spot. The collision energies in order of 12-24 GeV are higher enough from the energy of ionization of neutral atoms. Consequently, we can expect a new regime of collision ionization when femtosecond pulses propagate in gases. The main differentiation from the well-known multi-photon and tunnel ionizations, where all processes are included into the pulse width, and depend from intensity, the new ionization regime will realizes one conical emission of charged particles, due to the collisions between trapped particles and free ones.

## 7. Conclusions

Up to now, the basic experimental and theoretical investigations in the field are related to the study of radiation forces generated by laser beams and pulses acting on individual Rayleigh dielectric particles. In this paper the impact of the radiation force on continuous media (air) and an ensemble of neutral particles is explored. In the employed approach, the applied individual force to an atom is transformed to density force per volume. It is further considered that the optical response of dielectric media, connected with the propagation of laser pulses, is no stationary and nonlinear. As a result, analytical expression for the longitudinal density force and density potential of Gaussian laser pulse propagating in gaseous media is obtained. The recent theoretical and experimental results for diffraction-free regime [6] and nonlinear wave guiding propagation [7-9] of ultra-short femtosecond laser pulses, facilitate and enable the simplification and integration of these densities utilizing the first order of dispersion approximation in optics. Consequently, an effective longitudinal potential and force, acting in the media at level of the pulse width are obtained. The longitudinal radiation force depends on the time duration of the laser pulse. As the laser pulse duration decreases the force strength increases. In air the longitudinal potential of Gaussian laser pulse with energy of 1mJ is of twelve orders of magnitude greater than the Boltzmann energy of free particles. An important finding of this research is the demonstrated governed conditions under which the neutral particles can be confined in the pulse envelope and to move along with the group velocity. When the trapped particles into the pulse's spot are with high density, the probability of collision with the free atoms and molecules in air becomes significant. The collision energies are in 12 - 24 GeV range and are sufficient to ionize the neutral atoms. *Therefore, a new type of collision ionization can be observed, in condition of powerful femtosecond pulse propagation in air.* It is noteworthy to mention that the main difference of this new type of collision ionization from the multi-photon and tunnel ionizations, where all processes are included into the pulse width. *Thus, the new ionization regime will realizes one conical emission of charged particles, due to the collisions between trapped particles and free ones.*

Furthermore, the dipole interaction of the moving neutral particles with the electromagnetic field will be at the carrier to envelope frequency $\omega_{CEF} = k_0 (v_{ph} - v_{gr})$) instead the main ones $\omega_0$. This type of oscillation is in the sub-THz range for gases, THz in solids, and can be measured in a direction orthogonal to the direction of the laser pulse propagation. The dipole oscillation measured in the direction of propagation will be again with carrying frequency $\omega_0$ due to the Doppler effect. In nonlinear regime the neutral trapped particles will not generate at the third harmonics, but at a frequency proportional to three times carrier to envelope frequency $3\omega_{THz} = 3k_0 (v_{ph} - v_{gr})$. Such coherent sub THz (94 GHz) is experimentally observed in air [11, 12] and in different type glass plates [13-16] and explained theoretically by an avalanche parametric conversion mechanism [13].

## 8. Acknowledgments

The present work is funded by Bulgarian National Science Fund by grant No КП-06-ПН58/8-2021 and Bulgarian Ministry of Education and Sciences by grant ДО1-298-2021.